# Improved theory for the polarization-dependent transverse shift of a paraxial light beam in free space

Bekshaev A.

I.I. Mechnikov National University, Dvorianska 2, 65082, Odessa, Ukraine,
e-mail: bekshaev@onu.edu.ua

**Abstract**

Spatial distribution of the longitudinal field component of a circularly polarized optical beam depends on the polarization handedness, which causes the lateral shift of the beam "center of gravity" when its polarization toggles. We present the generalized theory of this effect, which demonstrates its relation with the angular irradiance moments of the beam. The theory is applicable to arbitrary paraxial beams and shows that the lateral shift is the same for all cross sections of the beam.

**Introduction**

Processes in which the polarization state affects spatial characteristics of a light beam are known as manifestations of the spin-orbital interaction of light [1–13]. Such effects can take place in presence of inhomogeneous or anisotropic media [2–8] as well as in free space [9–13]. One of the most spectacular manifestations of the spin-orbit coupling is the "spin Hall effect of light" – the polarization-induced transverse shift of the beam trajectory. Usually it is expressed by displacement of the "center of gravity" (CG) of the transverse energy distribution in the beam, depending on the handedness of its circular polarization [2–6,9–11,13]. In the classical version, this phenomenon is local and associates with the strong inhomogeneity occurring, e.g., at a plane boundary between different optical media [3–6]. However, the similar effect may take place in freely propagating optical fields (in particular, upon tight focusing of a perturbed Gaussian beam [11]). Then it cannot be attributed to a certain beam section and relates to the whole length of the beam [14] (is non-local).

Historically, the first effect of this sort was discovered in 1994 when the polarization-dependent shift of the focal spot in the $y$-direction was predicted [9] and observed [10] in conditions of focusing an initially collimated (paraxial) beam whose transverse profile was asymmetrical in the *x*-direction (Fig. 1). This shift was explained by employing the longitudinal (*z*-) component of the electric vector of the beam electromagnetic field whose spatial distribution shows explicit relation to the polarization handedness. However, the explanation given in [9] seems insufficient for two reasons:

(i) it essentially relies on the special model of the incident beam that was supposed in the form of the Hermite-Gaussian mode [15];

(ii) text of Ref. [9] creates the misleading impression as if the transverse shift is formed "on the passage" of light from the lens to the focal plane, and, when observing the focal plane image it cannot be seen because "at the stage of divergent wave propagation from the focal plane towards the microscope objective lens the effect of the opposite sign will identically nullify that shift" [9]. This assertion evidently contradicts to the known fact that the beam CG propagates along the geometric-optics trajectory, which is a straight line in the free space after the focusing lens.

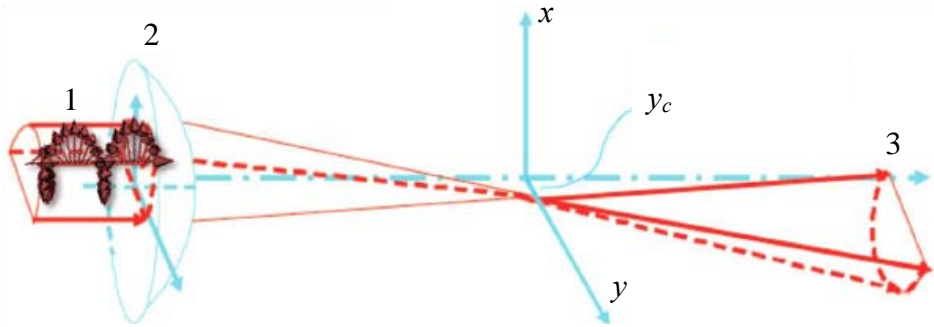

Fig. 1. Schematic of the experiment for detecting the polarization-dependent shift of the focal spot [10]: a circularly polarized beam 1 is focused by the lens 2; Cartesian frame (*x*, *y*) parameterizes the focal plane, the inverse image of the incident beam cross section is formed in the image plane 3. The asymmetry along the *x*-direction was created by screening the lower half of lens 2, the focal spot shift $y_c$ in the *y* direction is shown.

In fact, exhaustive description of the beam behavior in this situation must rely on the non-paraxial models that were recently developed basing on the plane-wave expansion of the light field [14,16]. However, following to Ref. [9], a consistent quantitative picture of the effect can be achieved by relatively simple means employing only paraxial concepts. In this note we present a generalization of the approach [9], which is applicable to arbitrary incident beam. Simultaneously, we intend to show that the same effect takes place in arbitrary cross section behind the lens, and that exclusive role of the focal plane is caused solely by the small transverse size of the focal spot. The absolute value of the transverse shift of the beam CG is everywhere the same but for the "narrow" focal spot the tiny effect can be seen easier than for any other cross section of the focused beam.

**Basic analysis**

Let us consider a monochromatic paraxial beam propagating along axis *z*. Its electromagnetic field can be represented as a superposition of *X*- and *Y*-polarized components with electric and magnetic vectors [15,17]

$$\begin{Bmatrix} \mathbf{E}_X \\ \mathbf{H}_X \end{Bmatrix} = \exp(ikz) \left( \begin{Bmatrix} \mathbf{e}_x \\ \mathbf{e}_y \end{Bmatrix} u_X + \frac{i}{k} \mathbf{e}_z \begin{Bmatrix} \partial/\partial x \\ \partial/\partial y \end{Bmatrix} u_X \right),$$

$$\begin{Bmatrix} \mathbf{E}_Y \\ \mathbf{H}_Y \end{Bmatrix} = \exp(ikz) \left( \begin{Bmatrix} \mathbf{e}_y \\ -\mathbf{e}_x \end{Bmatrix} u_Y + \frac{i}{k} \mathbf{e}_z \begin{Bmatrix} \partial/\partial y \\ -\partial/\partial x \end{Bmatrix} u_Y \right) \quad (1)$$

where $u_j$ ($j = X, Y$) are the slowly varying complex amplitudes. Longitudinal components of the field (1) are

$$E_z = \frac{i}{k}\left( \frac{\partial u_X}{\partial x} + \frac{\partial u_Y}{\partial y} \right), \quad H_z = \frac{i}{k}\left( -\frac{\partial u_Y}{\partial x} + \frac{\partial u_X}{\partial y} \right). \quad (2)$$

Within the frame of paraxial approximation, the longitudinal field (2) is small with respect to the transverse field, $E_z \sim \gamma(E_x, E_y)$, where the small parameter $\gamma$ coincides with the angle of self-diffraction (beam divergence) [17].

The energy density of the beam electromagnetic field can be represented in the form

$$w = \frac{1}{16\pi}\left( |\mathbf{E}|^2 + |\mathbf{H}|^2 \right) = w_\perp + w_z \quad (3)$$

where the first summand,

$$w_\perp = \frac{1}{8\pi}\left( |u_X|^2 + |u_Y|^2 \right), \quad (4)$$

owes to the transverse components of the field (1), and the contribution associated with the longitudinal field is

$$w_z = \frac{1}{16\pi}\left( |E_z|^2 + |H_z|^2 \right) \sim \gamma^2 w_\perp. \quad (5)$$

Our aim is to inspect the modification of the beam transverse position caused by the change of the incident beam polarization. This position is characterized by the CG of the transverse energy distribution

$$\mathbf{r}_c = \frac{\int \mathbf{r} w \, dx \, dy}{\int w \, dx \, dy} \quad (6)$$

where $\mathbf{r} = \begin{pmatrix} x \\ y \end{pmatrix}$ is the transverse radius-vector; from now on, all integrations are performed over the whole cross section of the beam. In agreement to (3), we can separate the contributions of $\mathbf{r}_c$ owing to the longitudinal and transverse field components: $\mathbf{r}_c = \mathbf{r}_{c\perp} + \mathbf{r}_{cz}$ with

$$\mathbf{r}_{c\perp} = \frac{\int \mathbf{r} w_\perp \, dx \, dy}{\int w \, dx \, dy} \approx \frac{\int \mathbf{r} w_\perp \, dx \, dy}{\int w_\perp \, dx \, dy} \tag{7}$$

and

$$\mathbf{r}_{cz} = \frac{\int \mathbf{r} w_z \, dx \, dy}{\int w \, dx \, dy} \approx \frac{\int \mathbf{r} w_z \, dx \, dy}{\int w_\perp \, dx \, dy}. \tag{8}$$

The second equalities in Eqs. (7) and (8) are possible due to small relative value of $w_z$, see Eqs. (3) and (5). Now note that in beams with homogeneous polarization, for which

$$u_Y = \beta u_X \tag{9}$$

with arbitrary complex constant $\beta$, the constituent $\mathbf{r}_{c\perp}$ (7) does not depend on polarization. This means that all the polarization influences are accumulated in expression (8). With allowance for (2) and (5) and due to small value of $w_z/w$ we can, in turn, represent it as a sum of two summands,

$$\mathbf{r}_{cz} = \mathbf{r}_{c1} + \mathbf{r}_{c2}, \tag{10}$$

where

$$\mathbf{r}_{c1} = \frac{1}{2k^2 I} \int \mathbf{r} \left( \left|\frac{\partial u_X}{\partial x}\right|^2 + \left|\frac{\partial u_X}{\partial y}\right|^2 + \left|\frac{\partial u_Y}{\partial x}\right|^2 + \left|\frac{\partial u_Y}{\partial y}\right|^2 \right) dx dy, \tag{11}$$

$$\mathbf{r}_{c2} = \frac{1}{2k^2 I} \int \mathbf{r} \left( \frac{\partial u_X^*}{\partial x} \frac{\partial u_Y}{\partial y} - \frac{\partial u_X^*}{\partial y} \frac{\partial u_Y}{\partial x} + \frac{\partial u_Y^*}{\partial y} \frac{\partial u_X}{\partial x} - \frac{\partial u_Y^*}{\partial x} \frac{\partial u_X}{\partial y} \right) dx dy \tag{12}$$

and

$$I = \int \left( |u_X|^2 + |u_Y|^2 \right) dx dy. \tag{13}$$

Again, under condition (9) the component (11) does not depend on $\beta$ and is thus invariant with respect to polarization. Therefore, it is the term (12) that expresses the sought spin Hall effect in the considered situation, and it will be in the focus of our further attention. Integration by parts, with taking into account that $u_X(x,y)$, $u_Y(x,y)$ and all their derivatives vanish at the transverse infinity, directly gives

$$\int x \frac{\partial u_X^*}{\partial x} \frac{\partial u_Y}{\partial y} dx dy = -\int \left( u_X^* \frac{\partial u_Y}{\partial y} + x u_X^* \frac{\partial^2 u_Y}{\partial x \partial y} \right) dx dy,$$

$$-\int x \frac{\partial u_X^*}{\partial y} \frac{\partial u_X}{\partial y} dx dy = \int x u_X^* \frac{\partial^2 u_Y}{\partial x \partial y} dx dy$$

so that

$$\int x \left( \frac{\partial u_X^*}{\partial x} \frac{\partial u_Y}{\partial y} - \frac{\partial u_X^*}{\partial y} \frac{\partial u_Y}{\partial x} \right) dx dy = -\int u_X^* \frac{\partial u_Y}{\partial y} dx dy.$$

Similarly,

$$\int x\left(\frac{\partial u_Y^*}{\partial y}\frac{\partial u_X}{\partial x} - \frac{\partial u_Y^*}{\partial x}\frac{\partial u_X}{\partial y}\right)dxdy = \int u_Y^* \frac{\partial u_X}{\partial y}dxdy,$$

whence we find the Cartesian component of vector $\mathbf{r}_{c2} = \begin{pmatrix} x_{c2} \\ y_{c2} \end{pmatrix}$,

$$x_{c2} = \frac{1}{2k^2 I}\int\left(u_Y^* \frac{\partial u_X}{\partial y} - u_X^* \frac{\partial u_Y}{\partial y}\right)dxdy. \quad (14)$$

In application to the *y*-component of (12), the same operations give

$$y_{c2} = \frac{1}{2k^2 I}\int\left(u_X^* \frac{\partial u_Y}{\partial x} - u_Y^* \frac{\partial u_X}{\partial x}\right)dxdy. \quad (15)$$

Introducing the circular components of the transverse field,

$$u_+ = \frac{1}{\sqrt{2}}(u_X - iu_Y), \quad u_- = \frac{1}{\sqrt{2}}(u_X + iu_Y),$$

one may represent (14) and (15) in the form

$$x_{c2} = \frac{i}{2k^2 I}\int\left(u_-^* \frac{\partial u_-}{\partial y} - u_+^* \frac{\partial u_+}{\partial y}\right)dxdy, \quad (16)$$

$$y_{c2} = \frac{i}{2k^2 I}\int\left(u_+^* \frac{\partial u_+}{\partial x} - u_-^* \frac{\partial u_-}{\partial x}\right)dxdy. \quad (17)$$

Eqs. (16) and (17) describe the spin Hall effect via the complex amplitude distributions of the circular polarization components of the beam under consideration. They can be represented in more clear form if we recall the expression for the first angular irradiance moments of the scalar beam with complex amplitude $u_\pm$ [18,19]

$$\begin{pmatrix} p_{x\pm} \\ p_{y\pm} \end{pmatrix} \equiv \mathbf{p}_\pm = -\frac{i}{kI_\pm}\int u_\pm^* \nabla u_\pm dxdy, \quad I_\pm = \int |u_\pm|^2 dxdy \quad (18)$$

(vector $\mathbf{p}_\pm$ expresses the mean value of the wavefront inclination with respect to the reference axis *z*, or, which is the same, the tilt of the CG trajectory in the given cross section). Then

$$x_{c2} = \frac{1}{2k}\left(p_{y+}\frac{I_+}{I} - p_{y-}\frac{I_-}{I}\right), \quad (19)$$

$$y_{c2} = \frac{1}{2k}\left(p_{x-}\frac{I_-}{I} - p_{x+}\frac{I_+}{I}\right). \quad (20)$$

**Discussion**

Eqs. (19) and (20) solve the maim problem of this paper as they describe the polarization-dependent part of the transverse position of the beam CG. They have transparent physical consequences; most important ones are the following:

- The transverse coordinates of the beam CG in any cross section depend on the spatial distribution of right and left polarization components of the incident beam, and vector $\mathbf{r}_{c2}$ inverts if the polarization handedness inverts in every point of the beam cross section.
- In beams which are superpositions of right and left polarized components, the net effect is determined by the difference between both contributions; each contribution enters with the weight proportional to the relative power of the corresponding component. E.g., for linear polarization, when these weights are equal, the effect disappears.
- Since vector $\mathbf{p}_\pm$ does not change in the course of free propagation, in all cross-sections the polarization-dependent shift of the beam CG should be the same; employing just the focal spot for its observation [10] is not principal although may be practical because only in the focal region the beam size is small enough to notice the tiny polarization shift.
- If the beam is polarized homogeneously (that is, condition (9) is true), $\mathbf{p}_+ = \mathbf{p}_- = \mathbf{p}$ and Eqs. (19), (20) reduce to

$$x_{c2} = \frac{\sigma}{2k} p_y, \quad y_{c2} = -\frac{\sigma}{2k} p_x \tag{21}$$

where $\sigma = (I_+ - I_-)/I$ is the (constant) polarization helicity. Note that the CG displacement $\mathbf{r}_{c2}$ is orthogonal to $\mathbf{p}$: $(\mathbf{p} \cdot \mathbf{r}_{c2}) = 0$. The case of Fig. 1 (the beam profile is symmetric with respect to axis $x$ but asymmetric with respect to axis $y$) corresponds to $p_y = 0$. Then, in application to the incident beam profile employed in Ref. [9], the first Eq. (18) and the second Eq. (21) yield the same result as the direct calculation of [9] based on explicit expressions of the complex amplitude distribution.

- Eqs. (19) and (20) also disclose the nature of the incident beam asymmetry that is necessary for the discussed effect: it reduces to requirement that $p_{x\pm}$ and/or $p_{y\pm}$ differ from zero. If the beam 'in itself' is symmetric (e.g., a circular Gaussian beam), this asymmetry can appear from the beam inclination (see Fig. 2). This situation reminds the spin-orbit and orbit-orbit effects associated with the oblique section of a light beam [13,20] but here, due to supposed paraxiality, the inclination angle $\theta$ is much less than the beam divergence – quite opposite to conditions accepted in [13,20].

In conditions of Fig. 2, the beam complex amplitude contains the phase factor $\exp(\mathrm{i}k\theta y)$, and according to (18), $p_y = \theta$. In agreement to (21), this entails

$$x_{c2} = \frac{\sigma}{2k}\theta \approx \frac{\sigma}{2k}\tan\theta. \tag{22}$$

This result remarkably simulates Eq. (16) of [13]; however, Eq. (22) deviates from the formulas of [13] in two important aspects. First, the CG definition in [13] differs from our definition (6) by replacement of the energy density by the longitudinal momentum density; second, in the geometry of Fig. 2 formulas of [13] predict the opposite sign of the CG displacement. The latter circumstance is not striking in view of the difference in CG definitions. However, the visual similarity of the analytical representations for both results is not occasional and reflects the universal geometric nature of the spin Hall effects of light in various systems [1].

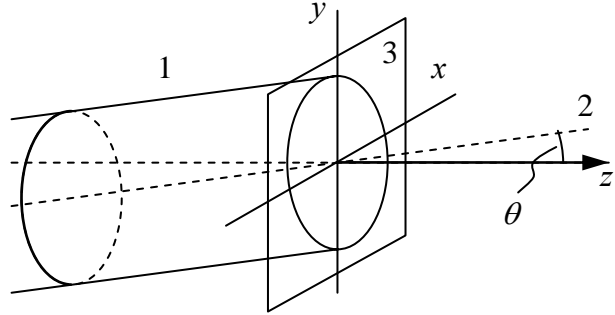

Fig. 2. Geometrical conditions in case where an inclined incident beam 1 with tilted propagation axis 2 is analyzed in the cross section 3 orthogonal to axis $z$ (other notations are explained in text)

**Conclusion**

To finalize the consideration, we emphasize that shift (19) – (21) of the beam CG (6) is completely caused by the polarization-dependent behavior of the longitudinal field (2). One could suppose that the focusing is essential for the effect because the longitudinal field components increase after the lens [21]. However, this is not the case, at least, as far as the focusing does not change the irradiance moments vector $\mathbf{p}_\pm$ (18). In general, a lens with the focal distance $f$ transforms this vector in accordance to relation [18,19]

$$\mathbf{p}_\pm = \mathbf{p}^i_\pm - \frac{\mathbf{r}^i_{c\pm}}{f} \qquad (23)$$

where superscripts "$i$" denote the quantities associated with ±-polarized components of the incident collimated beam. In Eq. (23), $\mathbf{r}^i_{c\pm}$ means the CG coordinates for the scalar beam model, i.e. only allows for the transverse field (zero-order of the paraxial approximation) and does not "include" the discussed effect. Eq. (23) says that if $\mathbf{r}^i_{c\pm} = 0$ (the "zero-order" CG of the incident beam is situated on the axis $z$), vector $\mathbf{p}_\pm$ does not change after focusing, and following to (19), (20), the discussed transverse shift of the focal spot is the same as the transverse shift of the incident beam CG. This is an additional confirmation that the use of just the focal spot for the transverse shift observation [9,10] is merely a technical expedient. Otherwise, the non-zero $\mathbf{r}^i_{c\pm}$ in Eq. (23) is equivalent to the incident beam asymmetry (cf. the half-lens screening in Fig. 1) and the corresponding term $-\mathbf{r}^i_{c\pm}/f$ expresses the focused beam inclination analogous to angle $\theta$ in Fig. 2.

The accuracy of the present analysis is limited by the paraxial approximation. In the light of the above discussion showing that the effect, in general, is not directly associated to the beam focusing, it seems quite appropriate. However, removal of the paraxial limitations will be useful in further development. In particular, it will allow considering how the analyzed shift (22) of the "energy" CG (6) interplays with other Hall effects which concern the longitudinal momentum distribution [13] or originate from the orbital angular momentum [20] of a beam.

**Acknowledgements**

This work was supported, in part, by the Ministry of Education and Science of Ukraine within the frame of the budget project No 457/09 (State Registration Number 0109U000942).

**References**

1. Bliokh K Y, Niv A, Kleiner V and Hasman E, 2008. Geometrodynamics of spinning light. Nature Photon. 2: 748–753.
2. Liberman V S and Zel'dovich B Y, 1992. Spin-orbit interaction of a photon in an inhomogeneous medium. Phys. Rev. A 46: 5199–5207.
3. Fedoseyev V G, 1991 Lateral displacement of the light beam at reflection and refraction. Opt. Spektrosk. 71: 829-834; Opt. Spektrosk. 71: 992–997.
4. Onoda M, Murakami S and Nagaosa N, 2004. Hall effect of light. Phys. Rev. Lett. 93: 083901.
5. Bliokh K Y and Bliokh Y P, 2007. Polarization, transverse shifts, and angular momentum conservation laws in partial reflection and refraction of an electromagnetic wave packet. Phys. Rev. E 75: 066609.
6. Bliokh K Y, 2009. Geometrodynamics of polarized light: Berry phase and spin Hall effect in a gradient-index medium. J. Opt. A: Pure Appl. Opt. 11: 094009.
7. Marrucci L, Manzo C and Paparo D, 2006. Optical spin-to-orbital angular momentum conversion in inhomogeneous anisotropic media. Phys. Rev. Lett. 96: 163905.
8. Fadeyeva T A, Rubass A F, and Volyar A V, 2009. Transverse shift of a higher-order paraxial vortex beam induced by a homogeneous anisotropic medium. Phys. Rev. A 79: 0538115.
9. Baranova N B, Savchenko A Y and Zel'dovich B Y, 1994. Transverse shift of a focal spot due to switching of the sign of circular polarization. JETP Lett. 59: 232–234.
10. Zel'dovich B Y, Kundikova N D and Rogacheva L F, 1994. Observed transverse shift of a focal spot upon a change in the sign of circular polarization. JETP Lett. 59: 766–769.

11. Volyar A and Fadeyeva T, 2000. Nonparaxial Gaussian beam: 2. Splitting of the singularity lines and the optical Magnus effect. Tech. Phys. Lett. 26: 740–743.
12. Zhao Y, Edgar J S, Jeffries G D M, McGloin D and Chiu D T, 2007. Spin-to-orbital angular momentum conversion in a strongly focused optical beam. Phys. Rev. Lett. 99: 073901.
13. Aiello A, Lindlein N, Marquardt C and Leuchs G, 2009. Transverse angular momentum and geometric spin Hall effect of light. Phys. Rev. Lett. 103: 100401.
14. Chun-Fang Li, 2008. Representation theory for vector electromagnetic beams. Phys. Rev. A 78: 063831.
15. Haus H. A. Waves and fields in optoelectronics. Englewood Cliffs, New Jersey: Prentice-Hall, Inc. (1984).
16. Bliokh K Y, Alonso M A, Ostrovskaya E A, and Aiello A, 2010. Angular momenta and spin-orbit interaction of non-paraxial light in free space. ArXiv:1006.3876v2 [physics.optics].
17. Bekshaev A Ya and Soskin M S, 2007. Transverse energy flows in vectorial fields of paraxial beams with singularities. Opt. Commun. 271: 332–348.
18. Anan'ev Yu A, Bekshaev A Ya, 1994. Theory of intensity moments for arbitrary light beams Opt. Spectr. 76: 558–568.
19. Mejias P M, Martinez-Herrero R, Piquero G and Movilla J M, 2002. Parametric characterization of the spatial structure of non-uniformly polarized laser beams. Prog. Quantum Electron. 26: 65–130.
20. Bekshaev A Ya, 2009. Oblique section of a paraxial light beam: criteria for azimuthal energy flow and orbital angular momentum. J. Opt. A: Pure Appl. Opt. 11: 094003.
21. Nieminen T A, Stilgoe A B, Heckenberg N R and Rubinsztein-Dunlop H, 2008. Angular momentum of a strongly focused Gaussian beam. J. Opt. A: Pure Appl. Opt. 10: 115005.